\documentclass[ams,prl,twocolumn,amssymb,superscriptaddress,longbibliography,floatfix]{revtex4-1}
\usepackage{amsmath}
\usepackage{tabularx}
\usepackage{bm}
\usepackage{slashed}
\usepackage{euscript}
\usepackage{graphicx}
\usepackage{xcolor}
\usepackage{amsfonts}
\usepackage{exscale}
\usepackage{amsbsy}
\usepackage{subfigure}
\usepackage{textcomp}
\usepackage{comment}
\usepackage{mathrsfs}
\usepackage{afterpage}

\usepackage[normalem]{ulem} %underlines OFF in hyperref

\usepackage{bbold}
\usepackage{blkarray}

\usepackage{ulem}

 \usepackage{hyperref}

\newcommand{\be}{\begin{equation}}
\newcommand{\ee}{\end{equation}}
\newcommand{\tr}{\textrm{tr}}

\newcommand{\beginsupplement}{%
   \setcounter{equation}{0}
   \pagebreak
\onecolumngrid
\setcounter{figure}{0}
\setcounter{table}{0}
\setcounter{page}{1}%
\renewcommand{\theequation}{S\arabic{equation}}%
\renewcommand{\thefigure}{S\arabic{figure}}
}

\begin{document}

\title{Interaction-induced metallicity in a two-dimensional disordered non-Fermi liquid} 

	\author{P. A. Nosov}
	
	\affiliation{Stanford Institute for Theoretical Physics, Stanford University, Stanford, California 94305, USA}
	
	\author{I. S. Burmistrov}
	\affiliation{L.D. Landau Institute for Theoretical Physics, acad. Semenova av.1-a, 142432, Chernogolovka, Russia}
	\affiliation{\hbox{Laboratory for Condensed Matter Physics, National Research University Higher School of Economics, 101000 Moscow, Russia}}
	
	\author{S. Raghu}
	
	\affiliation{Stanford Institute for Theoretical Physics, Stanford University, Stanford, California 94305, USA}
	
	\affiliation{\hbox{Stanford Institute for Materials and Energy Sciences, SLAC National Accelerator Laboratory, Menlo Park, CA 94025, USA}}
	\date{\today}
	
	\begin{abstract}
The interplay of interactions and disorder in two-dimensional (2D) electron systems has actively been studied for decades.  The paradigmatic approach involves starting with a clean Fermi liquid and perturbing the system with both disorder and interactions. We instead start with a clean non-Fermi liquid near a 2D ferromagnetic quantum critical point and consider the effects of disorder. In contrast with the disordered Fermi liquid, we find that our model does not suffer from runaway flows to strong coupling and the system has a marginally stable fixed point with perfect conduction.
	\end{abstract}

	\maketitle

	Despite enormous 
	progress \cite{Lee1985,Altshuler1985,Finkelstein1990,Belitz_Kirkpatrick_Rev1994,Castro2003,Kamenev2009,Finkelstein2010,Dobrosavljevi2010,dellAnna2017,Liao2017,Burmistrov2019}, 
	the possible ground states of 
	two dimensional (2D) interacting, disordered electron systems remain largely unexplored. 
 	In agreement with experimental observations \cite{2D_metals_review2001,Shangina2003,Kravchenko2003,Pudalov2004,Shashkin2005,Gantmakher2008,Kravchenko2010,Kravchenko2012,Kravchenko2017strongly,Dolgopolov2019,Kapitulnik2019},
	the theory of disordered Fermi liquids (FL) \cite{Finkelstein1984c,Castellani1984,FinkelStein_WeakLoc1984,Kirkpatrick1990a,Punnoose2001,Punnoose2005} suggests that in some cases, interactions can stabilize 2D metallic behaviour at low temperatures ($T$), while the non-interacting counterparts remain fully localized in 2D
	\cite{Abrahams1979}. However, the ultimate understanding of experiments requires well-controlled theories of strong interactions and disorder, representing a fundamental challenge.

	In the theory of 2D disordered FLs, metallic behavior occurs near a strong coupling fixed point, marking the onset of a magnetic instability \cite{FinkelStein_WeakLoc1984,Castellani1984}. This
instability has been interpreted as indicating either the formation of local moments \cite{FinkelStein_WeakLoc1984,Sachdev1989,Bhatt1992,Aleiner2000} or ferromagnetism \cite{Kirkpatrick1996,Ferromagnetism_Kamenev, New_Saddle_Point_NLSM_2000,Nayak2003}. At present, the strong coupling fixed point and the associated metallicity remain poorly understood. Experimental studies of 2D systems have revealed an enhancement of  electron spin susceptibility 
	\cite{Okamoto1999,Shashkin2001p,Vitkalov2001,Pudalov2002,Tutuc2002,Zhu2003,Pudalov2004,Clarke2007} and the existence of spin droplets \cite{Kuntsevich1,Kuntsevich2,Morgun2016,Pudalov2018} 
	in the metallic-like regime. Both theory and experiment call for an alternative  approach in which magnetic fluctuations are treated beyond a mean-field approximation.

Close to the ferromagnetic ordering, FL breaks down via scattering of fermions off soft magnetic fluctuations, leading to a `non-Fermi liquid' (NFL) \cite{Altshuler1995,Chubukov2006,Sung-Sik2018}. Within a phenomenological  approach, the 2D NFL with vanishing density of states (DOS) at the chemical potential is stable to localization and remains a perfect conductor \cite{Chakravarty1998}, in agreement with general scaling arguments \cite{Dobrosavljevi1997},
\footnote{Recently, a number of 2D NFL models with a vanishing DOS which exhibit dirty metallic behaviour
has been studied
\cite{QED3disorder2017, Goldman2017,QED3Thomson2017,Yerzhakov2018,Goldman2020,Mulligan2020}.}.
Much less is known about the effect of disorder on 2D NFLs having non-zero DOS \cite{Varma_Kondo_2002,Maslov2005}.

\begin{figure}[b]
\center{\includegraphics[width=0.8\linewidth]{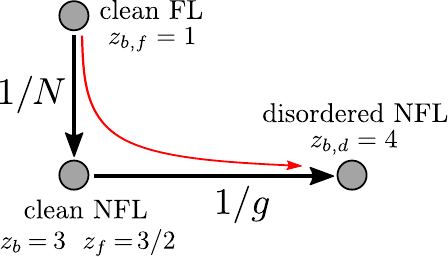}}
\caption{Schematic RG flow for a disordered 2D fermionic system interacting with the “magnetic” order parameter fluctuations. The intermediate clean NFL fixed point is unstable to disorder and ultimately leads to a dirty fixed point with the dynamical scaling $z=4$.}
\label{fig:RG_general}
\end{figure}

In this Letter, we study disorder effects near a  metallic quantum critical point %(QCP) 
at which singular effects of interactions lead to a magnetic instability. Since the strong interactions require additional control parameters, we start with a recently studied tractable large $N$ limit of a 2D NFL, which involves fermions  coupled to quantum critical ``magnetic" fluctuations \cite{matrixlargeN2019}.
Assuming that the characteristic energy scales of NFL behaviour and diffusion are well-separated, we incorporate the absence of quasiparticles already  at the saddle-point level, and study the combined effects of residual interactions and disorder by means of the renormalization group (RG) (see Fig.~\ref{fig:RG_general}).

%%%%%%%%%%%%%%%%%%%%%%%%%%%%%%%%%%
%FIGURE
%%%%%%%%%%%%%%%%%%%%%%%%%%%%

In contrast to the disordered FL, we find unconventional dynamical scaling of the diffusion propagator (diffuson) inherited from the NFL.  
Moreover,  short-range interactions are irrelevant in the RG sense in our theory, and runaway flows to strong coupling
disappear.  The only remaining source of infrared (IR) divergences is the small momentum scattering mediated by the diffusive Landau damped magnetic order parameter, 
which ultimately sets the new dynamical scaling $z_d=4$ for diffusons below a certain energy scale.  As a result, the coupling between diffusons and the magnetic fluctuations vanishes under the RG, giving way to a well-controlled fixed point with  perfect conduction.

%%%%%%%%%%%%%%%%%%%%%%%%%%%%%%%%%%%
\textbf{Summary of results.\,--} We consider a Euclidean Lagrangian
density
$\mathcal{L}=\mathcal{L}_0+\mathcal{L}_{\mathrm{int}}+\mathcal{L}_{\mathrm{dis}}$ which involves a %two-dimensional 
2D system %consisting 
of fermions $\psi$ at a finite density interacting with a critical magnetic collective mode $\phi$,
\be
\begin{aligned}\label{eq:model}
    \mathcal{L}_{0} &= \operatorname{tr}\left[ (\partial_\tau \phi)^2+(\nabla \phi)^2\right] + \bar{\psi}_j \left[\partial_\tau +\xi(i\nabla)\right]\psi^j \;,\\
    \mathcal{L}_{\mathrm{int}} &= \frac{\lambda}{\sqrt{N}}\phi^{j}_{\;k}\bar{\psi}_j \psi^{k}\;,\quad \mathcal{L}_{\mathrm{dis}} = \frac{1}{\sqrt{N}}V(x)\bar{\psi}_j \psi^{j}\;,
\end{aligned}
\ee
where $\xi(p)=p^2/2m -\mu$. Here $m$ denotes the fermion mass and $\mu$ is the chemical potential.
This model has a global $SU(N)$ flavor symmetry, 
with $\psi_i$ and $\phi^{i}_{\;j}$ transforming in the fundamental and  adjoint representations, respectively.
A random potential $V(x)$ coupled to the fermionic density has a Gaussian distribution with the zero mean and 
a variance $\langle V(x)V(x^\prime) \rangle = (2\pi \nu \tau)^{-1} \delta(x-x^\prime)$. Here $\nu$ is the DOS per  flavour and $\tau$ is the mean free time. We assume that $1/N$ and $1/g$ (where $g$ is the dimensionless Drude conductivity per  flavour measured in units $e^2/h$) are the only expansion parameters of the model. In order to distinguish the interaction-induced effects from localization corrections, we follow \cite{Finkelshtein1983} and consider a situation in which the Cooper channel is suppressed by a small time-reversal and parity breaking field.

The low-temperature behaviour of theory \eqref{eq:model}
is governed by the RG equations derived at the leading order in $1/N$ and $1/g$, with no restrictions on the strength of the Yukawa coupling
$\lambda$, cf. Eq. \eqref{eq:RG_equations_FP1}. 
They exhibit an IR marginally stable fixed point at
$1/g=\lambda=0$ with the following features:

\noindent $\bullet$ The large-$N$ low-energy quantum dynamics is set by $\sim N^2$ multiplet 
 diffusons with dynamical scaling $z_d=4$.

\noindent $\bullet$  The %local 
 average fermionic DOS
 $\nu(E)$
 diverges very weakly 
    at sufficiently small energies, $|E|\ll \Lambda_{4}$, 
    \be\label{eq:DOS}
   \nu(E)\simeq \nu_0 \exp \{\alpha\ln^2\left[\ln\left(\Lambda_{4}/|E| \right) \right] \}\;,
    \ee
where $\alpha \approx 0.104$ is  
the universal exponent and $\nu_0$ is the density of states per one flavour at the emergent energy scale $\Lambda_{4}$ below which the system flows to the fixed point. 

\noindent $\bullet$ The conductivity diverges as $T\rightarrow 0$, 
in a slow logarithmic manner with the universal exponent $s\approx 0.704$,
    \be\label{eq:conductivity}
    g(T) \simeq g_0^{1-s}\zeta_0^{s}\;\ln^s\left(\Lambda_{4}/T \right),\quad T\ll\Lambda_{4} .
    \ee
  Here $g_0$ and $\zeta_0$ are the conductance and dimensionless interaction strength 
  (see below) at the energy scale $\Lambda_{4}$.

\noindent $\bullet$ 
$N^2$ bosonic modes with dynamical  scaling $z_b=4$ are responsible for the anomalous temperature dependence of the specific heat  at $T\ll \Lambda_4$, $c_v\sim T^{2/z}$ with $z=4$.
    
\noindent We now turn to the detailed description of  these results. 

\textbf{An intermediate clean fixed point.\,-- } We begin with the UV limit, where both $\mathcal{L}_{\rm dis}$ and $\mathcal{L}_{\rm int}$ are relevant perturbations, so we are free to take into account first  the cubic
Yukawa interaction before introducing any effects of disorder. The large--$N$ solution in the clean limit stems from the coupled set of Schwinger--Dyson equations.
The vertex corrections can be neglected at large $N$.
Under these conditions, the clean large--$N$ solution immediately leads to two crucial effects    \cite{matrixlargeN2019}. At first, the boson self-energy is dominated by  the
Landau damping, 
$\Pi(\omega_n,q)=\gamma |\omega_n|/(N q)$. Here $q$ is the momentum, $\omega_n=2n \pi T$  
is the bosonic Matsubara frequency,
and
$\gamma =\nu \lambda^2/2v$ with $v=\sqrt{2m \mu}$.
 Secondly, fermions become dressed into a NFL, with a self-energy correction 
$\Sigma_f(\varepsilon_n)=i\beta N^{1/3} \left|\varepsilon_n\right|^{2 / 3} \operatorname{sgn}\varepsilon_n$ where $\beta \propto \lambda^{4/3}\mu^{-1/3}$ and  $\varepsilon_n=(2n+1)\pi T$ stands for fermionic Matsubara frequency.
$\Sigma_f(\varepsilon_n)$
is parametrically larger than the bare $i\varepsilon_n$ term at low energies. As a result, the clean interacting fixed point is described by the effective Lagrangian density $\mathcal{L}_{\rm eff}=\mathcal{L}_{f}+\mathcal{L}_{b}+\mathcal{L}_{\rm int}$ where
\begin{equation}
    \begin{aligned}
\mathcal{L}_{f} & = \bar{\psi}_{j,\varepsilon_n}\left[i\beta N^{1/3} \left|\varepsilon_n\right|^{2 / 3} \operatorname{sgn}\varepsilon_n-\xi(i\nabla)\right] \psi^j_{\varepsilon_n}\;, \\
\mathcal{L}_{b} & =
\operatorname{tr}\left[\phi_{\omega_n}(q)\left(q^{2}+\frac{\gamma}{N}\frac{\left|\omega_n\right|}{q}\right)\phi_{-\omega_n}(-q)\right] .
\label{eq:one_loop_clean}
\end{aligned}
\end{equation}
The dynamical exponents are $z_b=3$ for the boson, and $z_f=3/2$ for the fermion (where $z_f$ is defined with respect to the momentum component perpendicular to the Fermi surface). In addition, various symmetry-allowed  interactions, such as $\phi^4$ and four-Fermi forward scattering, become irrelevant, and only the Yukawa coupling $\lambda$ remains marginal. Further analysis beyond the planar limit reveals that there are no leading $1/N$ logarithmic corrections to \eqref{eq:one_loop_clean} which can potentially destabilize the fixed point at large but finite $N$ \cite{Altshuler1994}.

\begin{figure}[t]
\center{\includegraphics[width=0.95\linewidth]{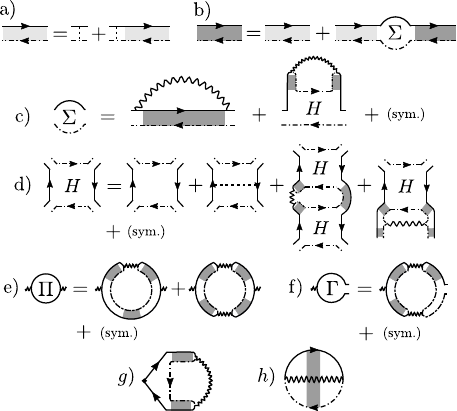}}
\caption{The shaded rectangular stands for a bare
diffuson (a). The impurity line is denoted by dashed line, and the retarded (advanced) single-particle
fermionic propagator is denoted by solid (dashed-dotted) line. Dark grey rectangular represents the diffuson (b) with self-consistent self-energy due to interaction shown in (c). Wavy solid line represents the dynamically screened bosonic propagator. White rectangular with symbol ‘$H$’ stands for the Hikami box which acquires self-consistent corrections (d). Leading in $1/N$ loop corrections to the bosonic self-energy are shown in (e) and the corrected boson-diffuson vertex is presented in (f). Representative diagrams for the tunneling density of states and the free energy are depicted in (g) and (h), respectively.}
\label{fig:leading_N_main}
\end{figure}

\textbf{Disorder at large $N$.\,--} Our next step is to re-introduce disorder $\mathcal{L}_{\rm dis}$ as a relevant perturbation at the one-loop Lagrangian \eqref{eq:one_loop_clean}. 
%From the diagrammatic perspective,
Diagrammatically,
we first dress the fermion propagator by non-crossing impurity lines within the self-consistent Born approximation,
%, such that 
%resulting in the form 
$G(\varepsilon_n,p)=[i(\beta N^{1/3} \left|\varepsilon_n\right|^{2 / 3} +(2N\tau)^{-1}) \operatorname{sgn}\varepsilon_n-\xi(p)]^{-1}$. 
%At  first glance, 
One might expect the disorder--induced lifetime to dominate at low energies,  rendering the NFL frequency dependence insignificant. However, the actual low-energy gapless degrees of freedom of the disordered  system are particle-hole excitations dressed with impurity ladders (diffusons), see Fig.\ref{fig:leading_N_main}a. 
%which in 
In our case they acquire anomalous dynamical scaling $z_d=3$ set by incoherent fermionic dynamics at low momenta, $q\ll (Nv\tau)^{-1}$, and frequencies, $ |\varepsilon_{n},\varepsilon_{n}^\prime|\ll (N^{4/3}\beta \tau)^{-3/2}$: $\mathcal{D}_{\varepsilon_n,\varepsilon_n^\prime}(q)= [NDq^2+N^{1/3}\beta (|\varepsilon_n|^{2/3}+|\varepsilon_n^\prime|^{2/3})]^{-1}$, where 
$D=v^2\tau/2$ is the diffusion constant. 
As a result, the diffusion pole in a particle-hole propagator (if analytically continued to real frequencies) is replaced by an effective IR-divergent energy relaxation time, which is, in some sense, very similar to the spin-flip time for triplet diffusons in the presence of dynamical magnetic impurities \cite{Ohkawa1983,Ohkawa1984,Vavilov2003,Kettemann2006,Micklitz2006,Micklitz2007,Kashuba2016,Burmistrov2018R}.

The fate of residual interactions can be now  naturally formulated in terms of anomalous diffusons. The appropriate field--theoretical description is given by the analog of the Finkel'stein nonlinear sigma model (FNLSM) which incorporates the non-trivial dynamical scaling already at the saddle point level. The details of this FNLSM approach are given in Supplemental Materials \cite{SM}. 
Below we derive the same results within the standard fermionic perturbation theory.

%%%%%%%%%%%%%%%%%%%%%%%%%%%%%%%%%%%%%%%%
%FIGURE
%%%%%%%%%%%%%%%%%%%%%%%%%%%%%%%%%

There are three major possible interaction effects, controlled by two small parameters, $1/N$ and $1/ g$. At first, the diffusons (which, by definition, include only processes with both particle and hole energies independently conserved) can be dressed by the self-energy effects via the Dyson equation (Fig.~\ref{fig:leading_N_main}b). Selection rules for both small parameters single out only diagrams of Fig.~\ref{fig:leading_N_main}c. Secondly, the bosonic propagator can also acquire corrections via the fermionic polarization operator. The simplest RPA-type resummation results in the form 
$(D^\phi_n(q))^{-1}= q^2+(2\beta)^{-1}\nu g^2|\omega_n|^{1/3}\;\mathcal{F}_{2/3}\left( Dq^2/\beta |\omega_n|^{2/3}\right)$ with the scaling function $\mathcal{F}_{\kappa}(x)= \int_{0}^{1}dt (x+t^{\kappa}+(1-t)^{\kappa})^{-1}$. 
At large $N$, the one-loop corrections to $D^\phi_n(q)$ are %significantly 
limited to %only 
two diagrams of Fig.~\ref{fig:leading_N_main}e only.
%due to large $N$. 
Finally, various vertex functions \cite{density_footnote} can also be renormalized by interactions. The only surviving diagram for quadratic boson-diffuson coupling is shown in Fig.~\ref{fig:leading_N_main}f, and the leading corrections to ``self-interactions'' between diffusons are depicted in Fig.~\ref{fig:leading_N_main}d.

\textbf{Self-consistent solution.\,--} The problem at hand very closely resembles the clean case, with diffusons playing a similar role that the fermions did in the clean limit.  
  We see this explicitly when we compute the one-loop diffuson self-energy diagrams (Fig.\ref{fig:leading_N_main}c) in a {\it self-consistent way} at the leading order in $1/(N
g)$, while ignoring higher order corrections (which turn out to be logarithmically divergent, so we will come back to them later). It is also worth noting that singular Hartree--type diagrams are absent because $\phi$ is traceless, and other processes involving large momentum transfer are either $1/N$ suppressed or irrelevant due to the anomalous dynamical scaling.

As in the clean case \cite{Sung-Sik2009}, the possibility to obtain a controllable self-consistent solution dramatically depends on how the large-$N$ and low-energy limits are simultaneously taken. We follow the procedure introduced in  \cite{matrixlargeN2019}, and rescale the bosonic and fermionic fields, momenta, and temperature as $\{\phi,\psi,\bar{\psi},q,T\}\rightarrow  \{N^2\phi,N^{\frac34}\psi,N^{\frac34}\bar{\psi},q/N,T/N^2\}$ with new %new 
$q,T\sim \mathcal{O}(N^0)$.
 Then the rescaled diffuson and bosonic propagator are free from any factors of $N$, and all $N$--dependence appears only in the vertices. We show
 \cite{SM} that within this rescaling none of the irrelevant operators are enhanced by a positive power of $N$. Physically, the rescaling procedure implies such a hierarchy of energy scales, when NFL effects (associated with $\gamma/N$) 
take place at energies higher than the onset of the diffusive regime (controlled by $1/(N \tau)$).

After the $N$--rescaling, the self-consistent solution for the diffuson propagator, see Fig.~\ref{fig:leading_N_main}b, takes the form
\be\label{eq:renormalized_diffuson}
%{eq:sigma_dif}
%\Sigma_{\varepsilon_n,\varepsilon_n^\prime}} 
(\mathcal{D}_{\varepsilon_n,\varepsilon_n^\prime}(q))^{-1}= Dq^2+\beta_{4}(|\varepsilon_n|^{1/2}+|\varepsilon_n^\prime|^{1/2})\, ,
\ee
where $\beta_{4}=\sqrt{(v\gamma/D)}/(4\pi\nu)$. Here the frequency dependence  $\sim|\varepsilon_n|^{1/2}$ comes from the self-energy correction to diffusion due to Yukawa interaction. This frequency dependence
overshadows the bare frequency dependence of the diffuson $\sim|\varepsilon_n|^{2/3}$ for energies below the emergent scale $\Lambda_{4}= (v\gamma/\beta^2D)^3$. In contrast, we did not find any non-analytic corrections to the bare momentum-dependence $\sim q^2$ {\it at the same} order $\mathcal{O}(1/\sqrt{g})$, and thus, the IR dynamical scaling of the diffuson becomes $z_d=4$.

The scaling \eqref{eq:renormalized_diffuson}  is ``self-consistent" in the following sense: %first, 
if one feeds this new diffuson back to the boson via quadratic vertices, then the dynamical scaling of the boson remains the same $z_b=4$,
\be\label{eq:renormalized_boson}
\bigl (D^\phi_n(q)\bigr )^{-1}=c^2 q^2+ \frac{v\gamma |\omega_n|^{1/2}}{\beta_{4}}\;\mathcal{F}_{1/2}\left( \frac{Dq^2}{\beta_{4} |\omega_n|^{1/2}}\right)\;,
\ee
where $\mathcal{F}_{1/2}(x)=2-\frac{\pi x}{2} +\frac{2(x^2-1)}{\sqrt{x^2-2}}\operatorname{arccot}\Big(\frac{2+x}{\sqrt{x^2-2}}\Big)$ and $c=1$. Although the exact form of the frequency-dependent term is modified, we find that asymptotically, for $q\gg \sqrt{\beta_{4}/D} |\omega_n|^{1/4}$, the standard Landau damping, 
$|\omega_n|/Dq^2$, restores. 
Then if we use the renormalized boson propogator, Eq.~\eqref{eq:renormalized_boson}, to re-evaluate the same self-energy diagram, Fig.~\ref{fig:leading_N_main}b, for the diffuson,  (that gave us $z_d=4$ in the first place), we obtain essentially the same result (up to small in $1/g$ corrections). Physically, the  Landau damping controls the dynamical scaling of appropriate low energy modes associated with fermions even in the dirty case (in the clean scaling $z_f=3/2$ is also dictated  by the ballistic form of the Landau damping).

The crucial observation here is that the coupling between the critical boson $\phi$ and the diffuson at the fixed point $z_{b,d}=4$ is {\it marginal}. Thus, the dirty and clean limits share the common strategy of finding a self-consistent solution that results from resumming the effects of interactions.   
Contrary to the clean case,  logarithmic divergences appear at $\mathcal{O}(1/g)$, and thus, require extra caution.

\textbf{Logarithmic corrections and RG.\,--} 
The scale-dependent corrections to the self-consistent solution should be derived by re-evaluating the diagrams in Fig.~\ref{fig:leading_N_main} with the modified propagators \eqref{eq:renormalized_diffuson} and \eqref{eq:renormalized_boson} near the fixed point $z_{b,d}=4$. There are no double-counting issues because the self-consistent propagators include only the leading correction $\sim \mathcal{O}(1/\sqrt{g})$, while logarithmic divergences $\sim\mathcal{O}(1/g)$ are sub-leading. The renormalization of the diffusion coefficient, the density of states and $\beta_{4}$ can be extracted from diagrams in Fig.~\ref{fig:leading_N_main}b-d. The divergent part of the bosonic self-energy diagrams depicted in Fig.~\ref{fig:leading_N_main}e renormalizes the coefficient $c^2$ %in front of the bare kinetic term $q^2$ 
in Eq. \eqref{eq:renormalized_boson}. The vertex correction of Fig.~\ref{fig:leading_N_main}f results in the renormalization of $\lambda$.

To extend the perturbative results into the RG form, 
we implement the following scaling procedure. We assign engineering dimensions as $[q]=1$, $[\bar\psi \psi]=-2+\eta_w$, $[\phi]=-2+\eta_\phi$, $[T]=4+\eta_T$, where $\eta_w$ and $\eta_\phi$ are anomalous field dimensions, and we also choose to scale temperature with some exponent $\eta_T$. This exponent is determined from the condition that the coefficient $\beta_{4}$ in the diffusion propagator \eqref{eq:renormalized_diffuson} does not run under the RG.

\begin{figure}[t]
\center{\includegraphics[width=0.95\linewidth]{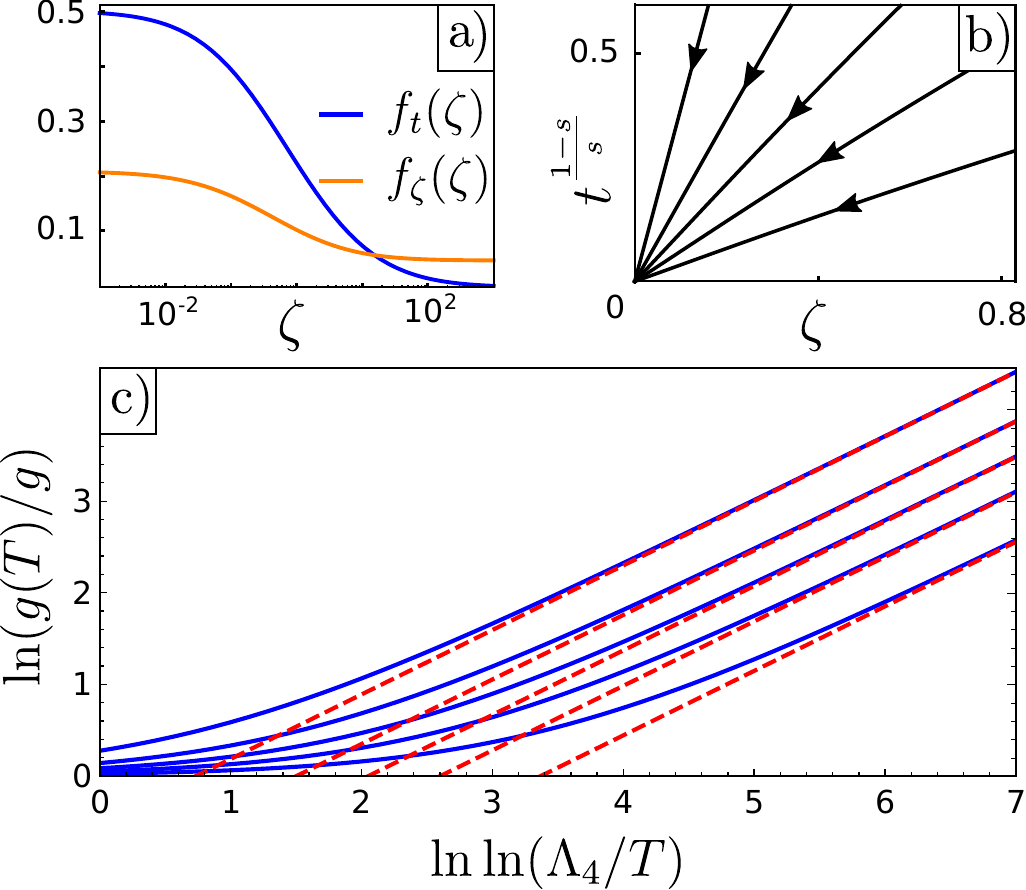}}
\caption{(a) The RG flow governed by Eqs.~\eqref{eq:RG_equations1:M}. The arrows mark direction of RG flow towards the infrared.
(b) The functions $f_t(\zeta)$ and $f_\zeta(\zeta)$ used in \eqref{eq:RG_equations1:M}.
(c) Temperature dependence of the logarithm of conductivity $\ln(g(T)/g)$ as a function of $\ln\ln(\Lambda_{4}/T)$ (blue curves). The red dashed lines represent an asymptotic behaviour \eqref{eq:conductivity}.  The used bare parameters are $t_0=0.3$ and $\zeta_0=0.16,\;0.35,\;0.60,\;1.04,\;2.25$ from bottom to top.}
\label{fig:RG_flow_lng}
\end{figure}

As a result, similar to the case of the disordered FL \cite{McMillan,Finkelshtein1983,Castellani1984}, the scaling of our theory is given by two-parameter RG flow for the dimensionless resistance $t=2/(\pi g)$ and the effective interaction $\zeta=(2\pi\beta_{4})^{-2}\lambda^2/t$, 
\begin{equation}
\label{eq:RG_equations1:M}
\frac{dt}{d\ln y} = - t^2 \zeta f_t(\zeta)\;, \quad
\frac{d\zeta}{ d\ln y} = - t \zeta^2 f_\zeta(\zeta)\;.
\end{equation}
Here $y$ is the running RG energy scale. The functions $f_{t,\zeta}(\zeta)$ are both positive, see Fig.~\ref{fig:RG_flow_lng}a, with the following asymptotic behavior: $f_t\approx (\ln \zeta)/(2\zeta)$, $f_\zeta\approx 0.048$ at $\zeta\gg 1$ and $f_t\approx 1/2+\zeta\ln\zeta$, $f_\zeta\approx \Delta-1/2$ at $\zeta\ll 1$ where $\Delta\approx 0.71$ \cite{SM}. We emphasize that we did not make any assumptions regarding the magnitude of Yukawa coupling $\lambda$, so that our RG equations are formally valid to all orders in $\lambda$. The RG flow governed by Eqs. \eqref{eq:RG_equations1:M} is depicted in Fig. \ref{fig:RG_flow_lng}b. Both the resistance $t$ and the normalized interaction $\zeta$ scale to zero values, improving the validity of our RG equations. There is only a single stable IR fixed point at $t=\zeta=0$. The asymptotic form of the RG equations near this fixed point is given by
\begin{equation}\label{eq:RG_equations_FP1}
\frac{1}{\Delta}\frac{dt}{d\ln y} = -  s \; t^2 \zeta\;, \qquad
\frac{1}{\Delta}\frac{d\zeta}{ d\ln y} = - (1-s) \; t \zeta^2\;,
\end{equation}
where $s=1/(2\Delta) \approx 0.704$. Eqs. \eqref{eq:RG_equations_FP1} indicate that both $t$ and $\zeta$ (and thus, also $\lambda$) are marginally irrelevant. One can immediately notice that all RG trajectories have the form $\zeta t^{(s-1)/s}=\text{const}$. By solving Eqs. \eqref{eq:RG_equations_FP1}, we find the scale dependence of the conductivity $g\sim 1/t$ depicted in Fig.~\ref{fig:RG_flow_lng}c, and with the asymptotic form given by Eq.~\eqref{eq:conductivity}.
From $\lambda^2 \sim t \zeta$ one can 
easily obtain the RG equation for the Yukawa coupling, $d\lambda^2 /d\ln y \sim - \lambda^4$. It yields $\lambda^2(E)\sim 1/\ln(\Lambda_{4}/E)$ in the infrared. We also note that all non-universal corrections to the anomalous dimensions vanish at this fixed point, $\eta_w,\eta_\phi,\eta_T \rightarrow 0$, implying that $z_d=4$ is a true dynamical scaling of the problem at low energies. 

%%%%%%%%%%%%%%%%%%%%%%%%%%%%%%%%%%
%FIGURE
%%%%%%%%%%%%%%%%%%%%%%%%%%%%

The correction to the tunneling DOS is shown in Fig.~\ref{fig:leading_N_main}g. As explained above, this perturbative correction can be cast in the form of RG flow. Near the fixed point the corresponding RG equation becomes \cite{SM},
$d\ln \nu/d\ln y=
t\zeta \ln(1/\zeta)/2$.
In the low-energy limit, $|E|\ll\Lambda_4$, this RG equation yields the asymptotic form given in Eq. \eqref{eq:DOS}. 

The behavior of the specific heat with $T$ can be deduced from interaction corrections to the free energy \cite{Castellani1986b}. Since the diagram of Fig.~\ref{fig:leading_N_main}h is IR finite, we find $c_v\sim T^{2/z}$ with $z=4$ \cite{SM}. This implies that, contrary to the case of disordered FL, the anomalous dimension $\eta_T$ does not influence the $T$--dependence of the specific heat.  
\color{black}

\textbf{Conclusions.\,--} We have developed the 
	RG theory describing the interplay between diffusive modes and soft ``magnetic" 
	fluctuations in a disordered 2D fermionic system. To keep the analysis parametrically under control, we employed expansion in two small parameters: $1/N$ (where $N$ corresponds to a global SU$(N)$ symmetry group) and dimensionless resistance $t$. By starting with an intermediate clean NFL fixed point, we found a self-consistent anomalous dynamical scaling $z_d=4$ of the multiplet
	diffusive particle-hole excitations (see Fig.~\ref{fig:RG_general}).  
	The residual infrared logarithmic corrections to this self-consistent solution were 
	summed up 
	by the RG equations, see Eq. \eqref{eq:RG_equations1:M}.
The RG flow exhibits an IR stable fixed point with both the resistivity and the residual interaction approaching zero in spite of weak divergence of the local density of states. 
By contrast, in the large $N$ extension of RG theory for the disordered FL \cite{Punnoose2005}, the magnetic instability still persists in the metallic regime.

While our theory predicts  a phase with perfect conduction when $t \ll 1$, 
we cannot exclude the existence of a metal--insulator transition expected on general grounds for sufficiently large disorder $t\sim 1$. Even in the non-interacting unitary case (without cooperons), certain higher order corrections due to diffusons favour suppression of conductivity \cite{WEGNER1989663} and can, in principle, compete with the interaction-induced effects \cite{Punnoose2005}.

Although it remains unknown whether our predictions hold in realistic systems with $N=2$, we may speculate on experimental implications of our results. 
We have provided  a model example where the resistance looks essentially finite for any realistically low accessible temperatures, but ultimately is zero at $T=0$. One can speculate that this behaviour is similar to a seemingly saturating resistance that has been observed in several experiments \cite{2D_metals_review2001} at low temperatures. In addition, recent experiments \cite{Kuntsevich1,Kuntsevich2,Morgun2016,Pudalov2018} indicate the formation of spin droplets near apparent metal-insulator transition, which makes the model studied in this paper to be extremely relevant for describing disordered 2D electron systems.

In the future, we wish to extend our results to the case when time-reversal symmetry is restored, leading to an  competition \cite{Tc_disorder} between weak-localization corrections and superconducting fluctuations. We expect both effects to be significantly modified because an anomalous dynamical scaling sets an unconventional temperature-dependence of the phase-breaking time, and generally enhances the BCS instability \cite{log2BCS2015}.

	\textbf{Acknowledgments.\,--} We thank A.~Chubukov, S.~Kachru, V.~Kravtsov for fruitful discussions. The work
of PAN and SR was supported in part by the US Department of Energy, Office of Basic Energy Sciences, Division of Materials Sciences and Engineering, under contract number DE-AC02-76SF00515. ISB acknowledges funding from the Russian Foundation for Basic Research (grant No. 20-52-12013) – Deutsche Forschungsgemeinschaft (grant No. EV 30/14-1) cooperation.
	
		\bibliography{NFLarxiv}

\begin{thebibliography}{80}
\bibitem [{den()}]{density_footnote}%
  \BibitemOpen
  \href@noop {} {}\bibinfo {note} {The singlet vertex is also strongly
  renormalized by multiplet diffusons, which effectively restore the diffusion
  pole in the density-density correlation function.}\BibitemShut {Stop}%
\end{thebibliography}

\afterpage{\null\newpage}

	\beginsupplement
	\begin{center}
  \textbf{\large  ONLINE SUPPORTING MATERIAL
  \\[.2cm] Interaction-induced metallicity in a two-dimensional disordered non-Fermi liquid
  }
\\[0.2cm]
P. A. Nosov$^{1}$, I. S. Burmistrov$^{2,3}$, and S. Raghu$^{1,4}$
\\[0.2cm]
{\small \it $^{1}$ Stanford Institute for Theoretical Physics, Stanford University, Stanford, California 94305, USA}

{\small \it $^{2}$ L.D. Landau Institute for Theoretical Physics, acad. Semenova av.1-a, 142432, Chernogolovka, Russia}

{\small \it \hbox{$^{3}$ Laboratory for Condensed Matter Physics, National Research University Higher School of Economics, 101000 Moscow, Russia}}

{\small \it \hbox{$^{4}$ Stanford Institute for Materials and Energy Sciences, SLAC National  Accelerator Laboratory, Menlo Park, CA 94025, USA}}

  \vspace{0.4cm}
  \parbox{0.85\textwidth}{In this Supplemental Material we (i) formulate the non-linear sigma model description for the considered system and derive  (ii) Eq. (5) of the main text, (iii) RG equations (7) of the main text, (iv) the RG equation for the tunneling density of states, and (v) scaling of the specific heat.} 
\end{center}

\section{Non-linear $\sigma$ model: definitions}\label{App_Sec:1}
In this section we discuss the $\sigma$-model formulation of the problem based on the one-loop action for the clean quantum critical point [Eq.(5) of the main text]. In order to capture the effect of disorder on this fixed point, we derive the non-linear $\sigma$ model (NL$\sigma$M) following the standard procedure: we average over disorder using the replica trick, decouple the resulting interaction via a matrix-valued Hubbard–Stratonovich field $Q$, and integrate out the fermions. The resulting integral over $Q$ is evaluated by the saddle-point method assuming all symmetry breaking terms to be small compared to $1/N\tau$. Finally, we obtain $\mathcal{S}=\mathcal{S}_{\sigma}+\mathcal{S}_b$ with
\be\label{eq:full_NLSM}
\mathcal{S}_{\sigma}=\frac{\pi\nu N}{4}\int d^2x \operatorname{Tr}\left[ D(\nabla Q)^2-4 N^{-2/3}\beta \;\hat{\eta}_{2/3} Q-4iN^{-3/2}\lambda T^{1/2} \hat{\phi} Q \right]\;,
\ee
where $D=v^2\tau/2$ is the diffusion constant, and the dynamical scaling is encoded into the matrix
\be
(\hat{\eta}_\delta)_{nn'}^{\alpha\beta,ij}=|\epsilon_n|^{\delta}\operatorname{sgn}\epsilon_n \;\delta_{nn'}\delta_{ij}\delta_{\alpha\beta}\;.
\ee
The matrix field $Q$ acts in the Matsubara (with index $n$ corresponding to Matsubara fermionic energies $\epsilon_n=\pi T (2n+1)$), replica (index $\alpha$) and $SU(N)$ (index $j$) spaces, and describes fluctuations around the replica-symmetric saddle point $\Lambda_{nn'}^{\alpha\beta ij}=\operatorname{sgn}(\epsilon_n) \delta_{nn'}\delta_{ij}\delta_{\alpha\beta}$. It also obeys the following constraints
\be 
Q=U^\dagger \Lambda U,\quad Q^2=1,\quad \operatorname{Tr}Q=0\;.
\ee
The last term in \eqref{eq:full_NLSM} describes the minimal coupling between the diffusons and the gapless boson $\phi$, which is now promoted to a matrix field acting in the full NL$\sigma$M space as $(\hat{\phi})_{nn'}^{\alpha\beta ij}= \phi_{n-n'}^{\alpha  ij}\delta_{\alpha\beta}$. Its action can be written as $\mathcal{S}_{b}=\int d^2x \operatorname{Tr}^\prime [\hat{\phi} \nabla^2 \hat{\phi} ]$, where $\operatorname{Tr}^\prime$ is normalized to the volume of the Matsubara space, and we dropped the bare frequency-dependent term because it will be generated dynamically.  

% The simple power-counting examination based on the fixed point action \eqref{eq:full_NLSM} shows that all higher-order vertices with two or more bosonic fields (as well as short-ranged fermionic interactions) are irrelevant due to the dynamical scaling $z_d=3$. 

For the perturbative expansion we shall use the square-root parametrization with the retarded-advanced blocks
\be \label{eq:W_parametrization}
Q=W+\Lambda \sqrt{1-W^{2}}, \quad W=\left(\begin{array}{cc}
{0} & {w} \\
{\bar{w}} & {0}
\end{array}\right)\,
\ee
where the fields $w$ and $\bar{w}$ satisfy $\bar{w}=w^\dagger$. Since $w$ is a complex matrix, we make a change of variables in a functional integral and consider $\bar{w}$ and $w$ as independent matrices. In addition, $\bar{w}$ and $w$ have the following matrix elements in the Matsubara space: 
 $\bar{w}_{n_1n_2}$ and $w_{n_2 n_1}$ are non-zero only for $n_1 <0$ and $n_2\geq 0$. We use the convention: $n_1,n_3,n_5,..<0$ and $n_2,n_4,n_6,..\geq 0$.

 To proceed further, we expand $Q=\sum_{n=1} Q^{(n)}$, where $n$ is the number of $W$ fields. The tree-level action $\mathcal{S}_{\rm Tree-level}$ is obtained by expanding \eqref{eq:full_NLSM} up to the second order in $W$ and $\phi$, with
\be\label{eq:tree_level_w}
\begin{aligned}
   \mathcal{S}_{\rm Tree-level}
    &=\frac{\pi\nu N}{4} \sum\limits_{\mathcal{A} \alpha\beta}\sum\limits_{n_1 n_2} \int_q \left[\bar{w}_\mathcal{A}(q)\right]_{n_1n_2}^{\alpha\beta}\left[Dq^2+N^{-2/3}\beta (|\epsilon_{n_1}|^{2/3}+|\epsilon_{n_2}|^{2/3}) \right]\left[w_\mathcal{A}(-q)\right]^{\beta\alpha}_{n_2n_1}\\
    &-\frac{i\pi \nu \lambda T^{1/2}}{2\sqrt{N}}\sum\limits_{nA\gamma}\int_q\; \phi_n^{\gamma A}(-q) \sum\limits_{n_1} \left([w_A(q)]^{\gamma \gamma}_{n_1-n,n_1}+[\bar{w}_A(q)]^{\gamma \gamma}_{n_1,n_1+n} \right) + \mathcal{S}_b\;,
    \end{aligned}
\ee
where $\int_{q} \equiv \int d^{2} q /(2 \pi)^{2}$, and all constraints on $n$-summations are implied according to our convention described above (i.e. $n_1+n$ and $n_1-n$ should be positive, etc.). We also decomposed both diffusion $W$ and bosonic $\phi$ fields as 
\be\label{eq:w_as_generators}
\phi_n^{\gamma}(q)=\sum\limits_{A=1}^{N^2-1}\phi_n^{\gamma A}(q) T^{A},\quad \left[\bar{w}(q)\right]_{n_{1} n_{2}}^{\alpha \beta} = \sum_{\mathcal{A}=0}^{N^2-1} \left[\bar{w}_\mathcal{A}(q)\right]_{n_{1} n_{2}}^{\alpha \beta} \tilde{T}^\mathcal{A}, \quad \left[w(q)\right]_{n_{1} n_{2}}^{\alpha \beta} = \sum_{\mathcal{A}=0}^{N^2-1} \left[w_\mathcal{A}(q)\right]_{n_{1} n_{2}}^{\alpha \beta} \tilde{T}^\mathcal{A}\;.
\ee
Here $\phi_n^{\gamma}(q)$ and $\left[\bar{w}(q)\right]_{n_{1} n_{2}}^{\alpha \beta}$ are viewed as matrices in the $SU(N)$ space. In addition, $\{\tilde{T}^\mathcal{A}\}$  ($\mathcal{A}=0,1,..,N^2-1$) is a complete set of matrices which consists of $SU(N)$ generators $\{T^A\}$ ($A=1,2,..,N^2-1$) and the normalized identity matrix. These matrices satisfy the following identities  
\be \label{eq:tilda_T}
\tilde{T}_{ij}^\mathcal{A}= \left\{\begin{array}{ll}
{T_{ij}^\mathcal{A}} & {\mathcal{A}\neq 0} \\
{(2N)^{-1/2}\delta_{ij}} & {\mathcal{A}=0}
\end{array}\right.,\quad\quad \sum\limits_\mathcal{A}\tilde{T}^{\mathcal{A}}_{ij} \tilde{T}^{\mathcal{A}}_{kl} =\frac{1}{2} \delta_{il} \delta_{kj},\quad \tr(\tilde{T}^\mathcal{A} \tilde{T}^\mathcal{B} )= \frac{1}{2} \delta^{\mathcal{A}\mathcal{B}}\;, \quad \tr(T^A \tilde{T}^\mathcal{B} )= \frac{1}{2} \delta^{A\mathcal{B}}\;.
\ee
The $SU(N)$ generators in the defining representation have several useful properties, such as
\be
\sum\limits_AT^{A\;j}_i T^{A\;l}_k =\frac{1}{2} \left(\delta_i^l \delta_k^j- \frac{1}{N} \delta_i^j \delta_k^k \right)\;, \quad \sum\limits_A T^AT^A=\frac{N^2-1}{2N}\;, \quad \sum\limits_A T^A T^B T^A = -\frac{1}{2N}T^B\;.
\ee

The majority ($\sim N^2$) of the degrees of freedom in \eqref{eq:tree_level_w} are multiplet diffusons with $\mathcal{A}\neq 0$. There is also a special mode with $\mathcal{A}=0$ (a singlet diffuson) corresponding to the density fluctuations. We note that due to the overwhelming number of multiplet diffusons, the singlet mode essentially "decouples" in the large-$N$ limit from the multiplet $\mathcal{A}\neq 0$ sector (consisting of multiplet diffusons and the matrix boson). More precisely, we show below that the singlet diffuson does not appear in any diagrams for the multiplet channel in the large-$N$ limit, allowing to solve the multiplet sector independently. However, there are leading self-energy corrections to the singlet propagator due to insertion of sub-diagrams involving multiplet diffusons, which modify the scaling behaviour of conductivity. In the last section, we show that a dynamical exponent for a singlet diffuson remains $z=2$, in full agreement with a Ward identity.

\subsection{Large-$N$ limit}
The tree-level terms for $W$ and $\phi$ fields explicitly contain factors of $N$ which makes it difficult to estimate how each particular diagram scales with $N$. Fortunately, it is possible to rescale the fields, momenta and temperature in such a way that all factors of $N$ will appear {\it only} in the vertices. Indeed, one can easily see that if we rescale
\be\label{eq:rescaling}
\quad q\rightarrow N^{-1} q, \quad T\rightarrow N^{-2}T,\quad W\rightarrow N^{3/2}W, \quad \phi \rightarrow N^2\phi\;,
\ee
with new $q,T\sim \mathcal{O}(N^0)$, then the tree-level terms $\mathcal{S}_{\rm Tree-level}$ are free from any $N$ factors. One can also check that \eqref{eq:rescaling} is consistent with the UV limits of all integrations/summations
\be\label{eq:UV_limits}
q\ll (v\tau N)^{-1}\; \rightarrow \; q\ll (v\tau)^{-1}, \quad\quad N^{1/3}\beta |\epsilon_n|^{2/3}\ll (N\tau)^{-1}\; \rightarrow \; \beta |\epsilon_n|^{2/3}\ll \tau^{-1}\;,
\ee
in a sense that all inequalities in \eqref{eq:UV_limits} now contain only $\mathcal{O}(N^0)$ quantities. Moreover, this procedure is consistent with the conditions for the Schwinger-Dyson equations in the clean limit. 
% In other words, the global $SU(N)$ symmetry implemented in our model allows to take the large-$N$ and low energy limits simultaneously in such a way that the anomalous dynamical scaling is preserved both for the boson and the diffuson. 

Note that the performed $N$-dependent rescaling \eqref{eq:rescaling} is not a scale transformation under which the tree-level terms are actually invariant. Indeed, such a transformation is given, for instance, by 
\be\label{eq:NLSM_scaling}
T\sim \lambda T, \quad q\sim \lambda^{1/3}q, \quad W/\phi \sim \lambda^{-2/3} W/\phi\;.
\ee
Thus, it is not expected for correlation functions to scale homogeneously to all orders under \eqref{eq:rescaling}. In particular, \eqref{eq:rescaling} also modifies the $N$-scaling of the interaction vertices. This is why the rescaling acts non-trivially over the action and makes the large-$N$ expansion more transparent. 

\begin{figure}[h!]
 \center{\includegraphics[width=0.7\linewidth]{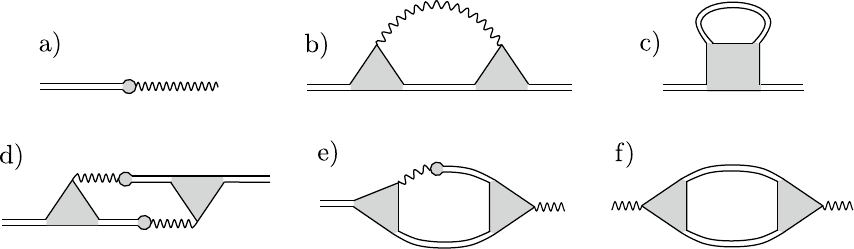}}
\caption{The skeleton diagrams in the multiplet sector $\mathcal{A}\neq 0$ contributing to the leading order in $1/N$ and $1/\tau$ after rescaling \eqref{eq:rescaling}. (a) The quadratic vertex mixing $W$ and $\phi$ fields. (b-d) Self-energy corrections to the multiplet diffuson. (e) The vertex correction. In the upper line, only the bare diffuson is assumed to prevent double counting. (f)
Correction to the bosonic propagator via the polarization operator.}
\label{fig:leading_N_diag}
\end{figure}

Now we check how the coupling with the boson (the last term in \eqref{eq:full_NLSM}) is affected by our rescaling \eqref{eq:rescaling}. After using the paramentrization \eqref{eq:W_parametrization}, we obtain the infinite number of vertices of the form $\operatorname{Tr}[\phi Q^{(n)}]$. By applying \eqref{eq:rescaling} we find, that these vertices scale with $N$ as $\sim N^{(1-n)/2}$. Therefore, only the linear in $W$ term (which is written explicitly in the second line in \eqref{eq:tree_level_w}) is $N$-independent. Diagrammatically, this term is represented by a gray circle connecting one double-line propagator for $W$ fields and the propagator for the boson $\phi$, as depicted in Fig.~\ref{fig:leading_N_diag}(a). The interaction terms quadratic in $W$ and linear in $\phi$ are depicted in Fig.~\ref{fig:leading_N_diag} as gray triangles. We emphasize that, as follows from \eqref{eq:NLSM_scaling}, the scaling dimension of the operator $\int dx \operatorname{Tr}[\hat{\phi} Q ]$ is $-1/6$, so the vertex is relevant. Similarly, one can compute the $N$-scaling of the next term in the expansion of $\mathcal{S}_\sigma$ (which is quartic in $W$ fields, and without the boson). This vertex yields $\sim N^{-1}$. From the perspective of the scaling transformation \eqref{eq:NLSM_scaling}, this vertex appears to be marginal.

 It is worth commenting on the other possible interaction terms that were left out of consideration in \eqref{eq:full_NLSM}. One possibility is a vertex containing $m$ bosons and $m$ matrices $Q$, each of which involves $n_j$ $W$-fields, $j=1,\dots, m$. The $N$-scaling of such vertex can be estimated as $N^{1-m/2-(n_1+n_2+...+n_m)}$. 
  Thus, this operator is not enhanced by any positive power of $N$. Moreover, its scaling dimension $m/2-2/3$ indicates that such vertices are irrelevant for $m\geq 2$. Another option is the short-ranged fermionic interactions, used in the standard Finkel'stein NL$\sigma$M. The typical vertex of this sort connects $m+m'$ diffusons and has the $N$-scaling $\sim N^{-(m+m')}$. The scaling dimension of this operator is always positive $+1/3$, so that the short-ranged interaction is irrelevant. This observation justifies why we ignored all such terms from the beginning.

In the coordinate representation, our $N$-rescaling \eqref{eq:rescaling} implies that the initial action \eqref{eq:full_NLSM} is transformed into
\be\label{eq:full_NLSM2}
\mathcal{S}_{\sigma}=\frac{\pi\nu N}{4}\int d^2x \operatorname{Tr}\left[ D(\nabla \tilde{Q})^2-4 \beta \;\widehat{\eta}_{2/3} \tilde{Q}-4i\lambda (T/N)^{1/2}\;\hat{\phi}\tilde{Q} \right]\;,
\ee
where the matrix field $\tilde{Q}=\frac{1}{\sqrt{N}}W + \Lambda \sqrt{1-\frac{1}{N}W^2}$ differs from $Q$ only by the normalization of $W$ fields. One can easily check that \eqref{eq:full_NLSM2} automatically reproduces correct $N$-factors for all vertices. The corresponding propagators for $\mathcal{A},\mathcal{B}$ components in \eqref{eq:full_NLSM2} can be represented as  
\be 
\begin{aligned}
  \label{eq:diffuson_plus_boson}
    &\left\langle\left[\bar{w}_\mathcal{A}(q)\right]_{n_{1} n_{2}}^{\alpha_{1} \beta_{1}}\left[w_\mathcal{A}(-q)\right]_{n_{4} n_{3}}^{\beta_{2} \alpha_{2}}\right\rangle = \frac{4}{\pi\nu }\mathcal{D}_{n_1n_2}(q) \Big( \delta_{n_1n_3}  -2\pi v\gamma T\mathcal{D}_{n_3n_4}(q)D^\phi_{n_1-n_2}(q) \delta_{\mathcal{A}\neq0}\delta_{\alpha_1\beta_1} \Big)\delta_{\alpha_1\alpha_2}\delta_{\beta_1\beta_2}\delta_{n_{12},n_{34}}\;,\\
    &(\mathcal{D}_{n_1n_2}(q))^{-1} = Dq^2+ \beta (|\epsilon_{n_1}|^{2/3}+|\epsilon_{n_2}|^{2/3})\;, \quad (D^{\phi}_n(q))^{-1}=q^2+\frac{\nu \lambda^2}{2\beta}|\omega_n|^{1/3}\;\mathcal{F}_{2/3}\left( \frac{Dq^2}{\beta |\omega_n|^{2/3}}\right)\;,
\end{aligned}
\ee
where $\langle \phi_n^{\gamma A}(q) \phi_{-n}^{\gamma A}(-q)\rangle =D^\phi_n(q)$ is a bosonic propagator, $\omega_n=2\pi T n$ is the bosonic Matsubara frequency, and $n_{ij}=n_{i}-n_{j}$. The scaling function $\mathcal{F}_{2/3}(x)$ is defined as $\mathcal{F}_{\kappa}(x)= \int_{0}^{1}dt ( x+t^{\kappa}+(1-t)^{\kappa})^{-1}$ with $ \mathcal{F}_{2/3}(0)\approx 0.83$. Note that $\langle w w\rangle$ and $\langle \bar{w} \bar{w}\rangle$ are zero. The asymptotic form of the dressed bosonic propagator depends on the characteristic momentum scale $\sqrt{\beta/D}|\omega_n|^{1/3}$. Specifically, if $q\gg\sqrt{\beta/D}\; |\omega_n|^{1/3}$ then  $(D_n^\phi(q))^{-1} \approx q^2+v\gamma\;\frac{|\omega_n|}{ Dq^2}$, and one recovers the standard diffusive Landau damping term with $z_b=4$.

\subsection{Perturbative expansion: interaction vertices}
So far we have discussed the tree-level terms (linear and quadratic in $W$ fields), and explained how to $N$-rescale all quantities. We now present the explicit form of the next terms in the expansion of $S_{\sigma}$. First, we find
\be 
\mathcal{S}^{(2,1)}_{\sigma}=-\frac{i\pi\nu \lambda T^{1/2}}{2\sqrt{N}} \sum\limits_{\substack{n\gamma A\mathcal{B}\mathcal{C} \\\beta n_1 n_2}} F^{A\mathcal{B}\mathcal{C}}\int\limits_{q_{1,2}} \phi_n^{\gamma A}(-q_1)  \left([\bar{w}_\mathcal{B}(q_2)]^{\gamma \beta}_{n_1-n,n_2}[w_\mathcal{C}(q_1-q_2)]^{ \beta \gamma}_{n_2,n_1}
 -
 [w_\mathcal{B}(q_2)]^{\gamma \beta}_{n_2-n,n_1}[\bar{w}_\mathcal{C}(q_1-q_2)]^{ \beta \gamma}_{n_1,n_2} \right)\;,
\ee
where $(n,m)$ stands for the vertex with $n$ diffusion $W$ fields and $m$ bosonic fields. Also, $F^{A\mathcal{B}\mathcal{C}} = \operatorname{tr}[T^A \tilde{T}^\mathcal{B}\tilde{T}^\mathcal{C}]$ is a structural constant. The general expression for $F^{A\mathcal{B}\mathcal{C}}$ is complicated, but we will need only several useful identities in the large--$N$ limit:
\be\label{eq:large_N_F}
\begin{aligned}
\sum\limits_{A\mathcal{B}}F^{A\mathcal{B}\mathcal{C}}F^{A\mathcal{C}\mathcal{B}} X_\mathcal{B} \rightarrow \frac{N}{8} X_{\neq 0}&\;,\quad \sum\limits_{A\mathcal{B}}F^{A\mathcal{B}\mathcal{C}}F^{A\mathcal{B}\mathcal{C}} X_\mathcal{B} \rightarrow \frac{N}{8} X_{\neq 0}\delta_{\mathcal{C}0}\;, \quad \sum\limits_{\mathcal{B}\mathcal{C}}F^{A\mathcal{B}\mathcal{C}}F^{A\mathcal{C}\mathcal{B}} X_\mathcal{B}X_\mathcal{C} \rightarrow \frac{N}{8} X_{\neq 0}^2\;,\\
&\sum\limits_{\mathcal{B}\mathcal{C}}F^{A\mathcal{B}\mathcal{C}}F^{A\mathcal{B}\mathcal{C}} X_\mathcal{B}X_\mathcal{C} \rightarrow 0\;, \quad \sum_{\mathcal{B}}F^{A\mathcal{B}\mathcal{B}}=0\;,
\end{aligned}
\ee
where $X_\mathcal{B}$ has only two values: $X_0$ and $X_{\neq 0}$. %, and we considered only the large-$N$ limit. 
Let us discuss the meaning of these relations. The first equation in \eqref{eq:large_N_F} suggests that the direct exchange diagram (with the internal multiplet $\mathcal{B}\neq 0$ diffuson) in Fig.~\ref{fig:leading_N_diag}(b) is the leading $1/N$ effect for any diffuson (with $\mathcal{A}=0$ or $\mathcal{A}\neq 0$). However, the second relation in \eqref{eq:large_N_F} indicates that the crossed exchange diagram (with the internal multiplet $\mathcal{B}\neq 0$ diffuson) also modifies the singlet diffuson $\mathcal{A}=0$ at the leading order in $1/N$. In fact, such crossed diagrams are responsible for the reemergence of the diffusion pole in the density-density correlation function. In addition, due to the last equation in \eqref{eq:large_N_F}, all 'Hartree-type' diagrams are exactly zero. The simple way to graphically understand these rules is to use a double-line representation for the bosonic propagator. 
% In this regard, the singlet $\mathcal{A}=0$ diffuson is special and can be viewed as a double line with closed ends, if used as an external element.
We emphasize that the singlet $\mathcal{A}=0$ diffuson never appears as an internal element due to the large-$N$ selection rules \eqref{eq:large_N_F}, and thus, is effectively decoupled from the multiplet sector, as was mentioned above.

Next, we consider the quartic vertex $S_{\sigma}^{(4,0)}$, which has the form
\begin{multline}
    S_\sigma^{(4,0)}=\frac{\pi\nu}{16N}\int\limits_{q_{1,..,4}}\delta\Big(\sum\limits_{i=1}^{4}\mathbf{q}_i\Big) \sum\limits_{\substack{n_1n_2\\n_3n_4}} \left[-D \;\Box_{q_1,q_2,q_3,q_4}+\beta (2\pi T)^{2/3}\sum\limits_{i=1}^{4}|n_i+1/2|^{2/3} \right] \sum\limits_{\mathcal{A}\mathcal{B}\mathcal{C}\mathcal{D}} J^{\mathcal{A}\mathcal{B}\mathcal{C}\mathcal{D}}\times
    \\
    \times\sum\limits_{\{\alpha_i\}}[\bar{w}_\mathcal{A}(q_1)]_{n_1n_2}^{\alpha_1\alpha_2}[w_\mathcal{B}(q_2)]_{n_2n_3}^{\alpha_2\alpha_3}[\bar{w}_\mathcal{C}(q_3)]_{n_3n_4}^{\alpha_3\alpha_4}[w_\mathcal{D}(q_4)]_{n_4n_1}^{\alpha_4\alpha_1}\;,
\end{multline}
 where we introduced the following notation
  \be
   \Box_{q_1,q_2,q_3,q_4} = 2(\mathbf{q}_1\mathbf{q}_3+\mathbf{q}_2\mathbf{q}_4)+(\mathbf{q}_1+\mathbf{q}_3)(\mathbf{q}_2+\mathbf{q}_4)\;,\quad
   J^{\mathcal{A}\mathcal{B}\mathcal{C}\mathcal{D}} = \operatorname{tr}\left[ \tilde{T}^\mathcal{A}\tilde{T}^\mathcal{B}\tilde{T}^\mathcal{C}\tilde{T}^\mathcal{D} \right]\;.
  \ee
The following identities will be useful: $\sum_{\mathcal{A}} J^{\mathcal{A}\mathcal{A}\mathcal{B}\mathcal{C}}X_\mathcal{A} \rightarrow \frac{N}{4} X_{\neq 0} \delta_{\mathcal{B}\mathcal{C}}$ and $\sum_{\mathcal{A}} J^{\mathcal{A}\mathcal{B}\mathcal{A}\mathcal{C}}X_\mathcal{A} \rightarrow \frac{N}{4} X_{\neq 0} \delta_{\mathcal{B}0}\delta_{\mathcal{C}0}$.

\section{Self-energy corrections to the multiplet $\mathcal{A}\neq 0$ diffuson
(Fig.~\ref{fig:leading_N_diag}(b-d))}\label{sec:SE_D}

Let us now consider the diagram in Fig.~\ref{fig:leading_N_diag}(b) in more details. We notice that due to the large-$N$ limit of the structural constants \eqref{eq:large_N_F} all relevant diagrams for the $\mathcal{A}\neq 0$ diffuson do not contain the singlet $\mathcal{A}= 0$ diffuson. Thus, one finds $\left[\Sigma^{(\rm{b})}_{\mathcal{A}\neq 0
}(q) \right]_{n_1n_2}^{n_3n_4}= \delta_{n_1n_3}\delta_{n_2n_4}\left[\Sigma^{(\rm{b},1)}_{\mathcal{A}\neq 0
}(q)\right]_{n_1n_2}^{n_1n_2}+\delta_{n_{12},n_{34}}\left[\Sigma^{(\rm{b},2)}_{\mathcal{A}\neq 0
}(q)\right]_{n_1n_2}^{n_3n_4}$, 
where the diagonal and off-diagonal parts read as
\begin{align}\label{eq:Sigma_1_1}
    \left[\Sigma^{(\rm{b},1)}_{\mathcal{A}\neq 0
}(q)\right]_{n_1n_2}^{n_1n_2}&=-\frac{\pi v \gamma}{4} T\sum\limits_{m_2}\int_k D^\phi_{m_2-n_2}(k)\mathcal{D}_{n_1m_2}(k+q) + (...)_{\substack{n_2\rightarrow -n_1 \\n_1\rightarrow -n_2} }\;,\\
\label{eq:Sigma_1_2}
 \left[\Sigma^{(\rm{b},2)}_{\mathcal{A}\neq 0
}(q)\right]_{n_1n_2}^{n_3n_4}&=\frac{(\pi v \gamma)^2}{2} T^2\hspace{-1em}\sum\limits_{\substack{ m_2\\m_2\geq n_2-n_4}}\hspace{-1em}\int_k D^\phi_{m_2-n_2}(k+q)D^\phi_{m_2-n_1}(k)\mathcal{D}_{n_1,m_2}(k)\mathcal{D}_{n_3,m_2+n_4-n_2}(k) + (...)_{\substack{n_{1,2}\rightarrow -n_{2,1} \\n_{3,4}\rightarrow -n_{4,3}}}\;.
\end{align}
From the scaling arguments, one can see that \eqref{eq:Sigma_1_2} generates a qudratic coupling of $W$ fields, with a structural factor scaling as the bosonic propagator. Thus, it plays a role similar to that of the bare linear vertex. However, due to the internal integration over momenta $q$, this vertex is at least $1/D$ suppressed, and can be neglected compared to the bare vertex which does not contain such small factors. The same reasoning holds for the next diagram depicted in Fig.~\ref{fig:leading_N_diag}(d)
\be
\left[\Sigma^{(\rm{d})}_{\mathcal{A}\neq 0
}(q)\right]_{n_1n_2}^{n_3n_4}=-\frac{(\pi v \gamma)^2}{2}\delta_{n_{12},n_{34}} T^2\hspace{-1em}\sum\limits_{\substack{ m_2\\m_2\leq n_2-n_4}}\hspace{-1em}\int_k D^\phi_{m_2-n_2}(k+q)D^\phi_{m_2-n_1}(k)\mathcal{D}_{n_1,m_2}(k)\mathcal{D}_{m_2+n_4-n_2,n_4}(k) +(...)_{\substack{n_{1,2}\rightarrow -n_{2,1} \\n_{3,4}\rightarrow -n_{4,3}}}\;,
\ee
Moreover, this diagram vanishes if we set external frequencies to zero $n_i\rightarrow 0$ due to the limited range of integrations.

Next, we investigate Fig.~\ref{fig:leading_N_diag}(c)
\be\label{eq:sigma_c}
\left[\Sigma^{(c)}_{\mathcal{A}\neq 0
}(q)\right]_{n_1n_2}^{n_1n_2}=\frac{\pi v\gamma}{4} 
\;T\sum\limits_{m_2}\int_k \Big\{[\mathcal{D}_{n_1n_2}(q)]^{-1}+[\mathcal{D}_{m_2n_2}(k)]^{-1} \Big\}\mathcal{D}_{m_2n_2}^2(k) D^\phi_{m_2+n_2}(k) + (...)_{\substack{n_2\rightarrow -n_1\\n_1\rightarrow -n_2}}
\ee
% \\+
% \frac{\pi v\gamma}{4}  \;T\sum\limits_{m_2}\int\frac{d^2k}{(2\pi)^2} \Big\{[\mathcal{D}_{n_1n_2}(q)]^{-1}+[\mathcal{D}_{n_1m_2}(k)]^{-1} \Big\}\mathcal{D}_{n_1m_2}^2(k) D^\phi_{m_2-n_1}(k)\;,
We will denote the first term in brackets in \eqref{eq:sigma_c} as $\Sigma^{(c,1)}$, and the second term as $\Sigma^{(c,2)}$. For now, let us put $\Sigma^{(c,1)}$ aside because it is trivially proportional to the inverse diffuson. Instead, we combine $\Sigma^{(b,1)}$ and $\Sigma^{(c,2)}$ together. One can easily see that in the limit $q=0$ and $n_1=n_2=0$ these diagrams cancel each other. Thus, we need to expand them in powers of small external momenta and frequency. Let us first set $q=0$, $n_1=-n_2$, and expand in small $n_2$
\be\label{eq:sigma_ce}
 \left[\Sigma^{(b,1)}_{\mathcal{A}\neq 0
}(0)+\Sigma^{(c,2)}_{\mathcal{A}\neq 0
}(0)\right]_{-n_2,n_2}^{-n_2,n_2}\approx \frac{\pi v \gamma}{2}\int\frac{d\epsilon_{m}}{(2\pi)}\int\frac{d^2k}{(2\pi)^2} D^\phi_{m}(k)\Big\{\theta(\epsilon_m-\epsilon_{n_2})\mathcal{D}_{m-n_2,n_2}(k)
-\theta(\epsilon_m+\epsilon_{n_2})\mathcal{D}_{m+n_2,n_2}(k)\Big\}\;.
\ee
As a result, we obtain
\be\label{eq:sigma_full_z_4}
\left[\Sigma^{(b,1)}_{\mathcal{A}\neq 0
}(0)+\Sigma^{(c,2)}_{\mathcal{A}\neq 0
}(0)\right]_{-n_2,n_2}^{-n_2,n_2} \approx  -\frac{1}{8}\sqrt{\frac{v\gamma}{D}}|\epsilon_{n_2}|^{1/2} +\mathcal{O}\Big(|\epsilon_{n_2}|^{2/3}\log (|\epsilon_{n_2}|)\Big)\;.
\ee
Note that the pre-factor is formally of the order $\mathcal{O}(1/\sqrt{\tau})$, and not $\mathcal{O}(1/\tau)$, because the bare bosonic kinetic term $q^2$ does not contain the difussion constant $D$. The emergent scale associated with the new dynamical regime is given by 
\be\label{eq:lambda_z_4}
\Lambda_{4}\sim \left(\frac{v\gamma}{\beta^2D}\right)^3\;.
\ee

 We note that similar calculations can be performed in order to obtain the leading momentum-dependent part of the self-energy. However, as can be easily seen, the expansion in powers of $Dq^2$ is regular and leads only to corrections to the diffusion coefficient. These corrections should be cut-off at the scale $\Lambda_{4}$. We emphasize that the contribution originating from the scales below $\Lambda_{4}$ should be computed with the new diffusion propagator which includes the self-energy effect \eqref{eq:sigma_full_z_4}. We will come back to such corrections in the next section.

By solving the Dyson equation, we incorporate \eqref{eq:sigma_full_z_4} into the diffusion propagator and set the UV cut-off of our theory to $\Lambda_{4}$ \eqref{eq:lambda_z_4}. The diffusion propagator is now given by
\be\label{eq:D_z=4}
\mathcal{D}_{n_1n_2}(q) = \frac{1}{Dq^2 + \beta_{4}(|\epsilon_{n_1}|^{1/2}+|\epsilon_{n_2}|^{1/2})}\;,\quad \beta_{4} = \frac{1}{4\pi\nu}\sqrt{\frac{v\gamma}{D}}\;.
\ee

Consequently, the bosonic propagator is also decorated with the new diffuson \eqref{eq:D_z=4} via quadratic mixing vertices
\begin{align}
 \label{eq:Pi_z=4_new}
(D^{\phi}_n(q))^{-1}&=  c^2q^2+v\gamma\beta_{4}^{-1}\;|\omega_n|^{1/2}\;\mathcal{F}_{1/2}\left( \frac{Dq^2}{\beta_{4} |\omega_n|^{1/2}}\right)\;,\\
\mathcal{F}_{1/2}(x)&=\int\limits_{0}^{1} \frac{dt}{x+t^{1/2}+(1-t)^{1/2}}=2-\frac{\pi x}{2} +\frac{2(x^2-1)}{\sqrt{x^2-2}}\operatorname{arccot}\left(\frac{2+x}{\sqrt{x^2-2}}\right)\;,
\end{align}
where $c^2=1$, $\mathcal{F}_{1/2}(0)=2-\sqrt{2}\operatorname{arccoth}(\sqrt{2}) \approx 0.753$, and $\mathcal{F}_{1/2}(x\gg 1)\approx 1/x$.

It is natural to ask if $z_d=4$ dictated by \eqref{eq:D_z=4} is a self-consistent dynamical scaling. To show that, we re-compute the same self-energy diagrams, but using \eqref{eq:D_z=4} and \eqref{eq:Pi_z=4_new} instead of the bare propagators \eqref{eq:diffuson_plus_boson}. This leads to the same answer dominated by the Landau damping term (up to IR-finite small corrections to $\beta_{1/2}$ proportional to $D^{-1}\log{D})$). Note that the Hikami box (a quartic vertex $S_\sigma^{(4,0)}$) is also self-consistently modified with the new operator $\sim \hat{\eta}_{1/2}$. This vertex ultimately results in logarithmic corrections to $\beta_{4}$ (originating from the analog of $\Sigma^{(c,1)}$). However, such corrections contribute to the order $\mathcal{O}(1/\tau)$, and not to $\mathcal{O}(1/\sqrt{\tau})$. Therefore, we can conclude that \eqref{eq:D_z=4} is indeed a self-consistent solution at the order $\mathcal{O}(1/\sqrt{\tau})$ with the dynamical scaling $z_d=4$. This self-consistent solution is subjected to logarithmic corrections at the order $\mathcal{O}(1/\tau)$, which we take into account in the next section by means of the RG.

\section{One-loop renormalization of the $\sigma$-model near the fixed point $z_d=4$}\label{App_Sec:2}
The remaining one-loop logarithmic corrections should be calculated with the modified propagators \eqref{eq:D_z=4} and \eqref{eq:Pi_z=4_new}. 
The corresponding fixed point action for the $\tilde{Q}$ matrices can be written as (cf. with \eqref{eq:full_NLSM2})
\be\label{eq:NLSM_z=4}
\mathcal{S}_{ z_d=4}=\frac{\pi\nu N}{4}\int d^2x \operatorname{Tr}\left[ D(\nabla \tilde{Q})^2-4 \beta_{4} \;\widehat{\eta}_{1/2} \tilde{Q}-4i\lambda (T/N)^{1/2}\;\hat{\phi}\tilde{Q} \right]\;,
\ee
where $\tilde{Q} = \frac{1}{\sqrt{N}}W + \Lambda \sqrt{1-\frac{1}{N}W^2}$. This fixed point with $z_{b,d}=4$ implies that the coupling constants ($D$ and $\lambda$) are marginal at the tree-level, allowing for the RG resummation. To find the fluctuation corrections to their scaling behavior,
we employ a Wilsonian RG scheme in which we integrate out energy shells from $y^{-1}\Lambda$ to $\Lambda$, where $y>1$ is the running RG scale, and $\Lambda$ is the UV cutoff. After each step of integration, we absorb emerging log divergences into the renormalization of the couplings according to the following scaling transformations
\be
q\rightarrow y^{-1} q, \quad W\rightarrow y^{2+\eta_w} W, \quad \phi\rightarrow y^{2+\eta_\phi}\phi, \quad T\rightarrow y^{-4+\eta_T}T\;,
\ee
where $\eta_w$ and $\eta_\phi$ are anomalous field dimensions. Here we choose to scale temperature with some non-trivial dimension $\eta_T$, which is determined from the condition that the coefficient $\beta_{4}$ in front of the $|\epsilon_{n_{1,2}}|^{1/2}$-term in \eqref{eq:D_z=4} does not run under the RG. It may appear that our RG scheme differs from the conventional scaling procedure introduced by Finkel'stein (which treats the coefficient $z_F$ in front of the frequency-dependent term $|\omega|$ as a coupling constant). However, both schemes, in fact, lead to the same RG equations if one uses a proper change of variables such that the RG equation for $\beta_{4}$ (or $z_F$ in the FL case) decouples from the rest of the system.

\subsection{Scale-dependent corrections to the diffuson}
We first compute corrections to the one-point correlation function $\nu(\epsilon_n)=\frac{\nu}{N}\langle \operatorname{Tr}[Q^{\alpha \alpha}_{nn}(x)]\rangle$ which represents the tunneling density of states per one fermion flavour. We find it to be log-divergent as 
\be
\frac{\delta\nu}{\nu}= t \zeta f_0(\zeta)\ln y\;,\quad f_0(\zeta)=\frac{1}{2}\int\limits_0^{+\infty}\frac{dy}{(y+1)^2(y+\zeta \mathcal{F}_{1/2}(y))}\approx\begin{cases}
\frac{1}{2}\ln\left(\frac{1}{\zeta}\right),\quad \zeta\ll 1\;\\
\frac{1}{4\zeta}\ln\left(1+\zeta\right),\quad \zeta \gg 1
\end{cases}
\ee
 Here we defined the normalized interaction strength $\zeta=(2\pi\beta_{4})^{-2}\lambda^2/t$. The anomalous dimension $\eta_w$ is determined to absorb the logarithmic divergence in $\langle \operatorname{Tr}[Q^{\alpha \alpha}_{nn}(x)]\rangle$. Thus, we obtain
\be\label{eq:etaW}
d\ln \nu/d\ln y=
t\zeta f_0(\zeta) \;,\quad \eta_w=t\zeta f_0(\zeta)\;.
\ee

The full self-energy for the diffuson is obtained by computing the same diagrams from Fig.~\ref{fig:leading_N_diag} with the propagators \eqref{eq:D_z=4} and \eqref{eq:Pi_z=4_new}. The correction to the momentum dependence of the self-energy reads as
\be\label{eq:sigma_full_z_4_qq}
\left[\Sigma^{}_{\mathcal{A}\neq 0
}(q)\right]_{00}^{00}= \frac{\pi v\gamma}{2}\Big(\int d\epsilon_m \int_k \mathcal{D}_m^2(k)D_m^\phi(k)- \int d\epsilon_m\int_k (Dk^2)\mathcal{D}_m^3(k)D_m^\phi(k) \Big) Dq^2\;.
\ee
By combining $[\Sigma_{\mathcal{A}\neq 0}^{(c,1)}(0)]_{n_1n_2}^{n_1n_2}$ (the first term in \eqref{eq:sigma_c}) together with \eqref{eq:sigma_full_z_4_qq}, we obtain
\be \label{eq:Sigma_FULL}
\begin{aligned}
\left[\Sigma_{\mathcal{A}\neq 0
}(q)\right]_{n_1n_2}^{n_1n_2} &=\Big(\Sigma^{(1)}  Dq^2  + \Sigma^{(2)} \beta_{1/2} (|\epsilon_{n_1}|^{1/2}+|\epsilon_{n_2}|^{1/2})\Big)\ln y\;,\\
\Sigma^{(1)} &=t\zeta\left(2f_0(\zeta)-f_{1}(\zeta)\right),\quad \Sigma^{(2)}=t\zeta f_{0}(\zeta)\;.
\end{aligned}
\ee
The positive and monotonic function $f_1(\zeta)$ is defined as
\be
f_1(\zeta)=\int\limits_0^{+\infty}dy\frac{y}{(y+1)^3(y+\zeta \mathcal{F}_{1/2}(y))}\approx\begin{cases}
\frac{1}{2}-\zeta\ln\left(\frac{1}{\zeta}\right),\quad \zeta\ll 1\;\\
\frac{1}{2\zeta}\ln\left(1+\zeta\right),\quad \zeta \gg 1
\end{cases}\;
\ee
By inspecting the self-energy \eqref{eq:Sigma_FULL} we can derive the following corrections to the diffusion constant and $\beta_4$
\begin{align}\label{eq:corrected_D_beta}
    \frac{\tilde{D}}{D} &= e^{2\eta_w \ln y} \left(1-t\zeta (2f_0(\zeta)-f_1(\zeta))\ln y\right)=1+ t \zeta f_1(\zeta)\ln y\;,\\
    \frac{\tilde{\beta}_{4}}{\beta_{4}} &=e^{(2\eta_w+\eta_T/2) \ln y} \left(1-t\zeta f_0(\zeta)\ln y\right)= 1+(\eta_T/2+ t\zeta f_0(\zeta))\ln y\;.
\end{align}
where $\tilde{D}$ and $\tilde{\beta}_{4}$ are modified couplings. By differentiating \eqref{eq:corrected_D_beta} with respect to $\ln y$, we obtain the RG equation for $t$, and the expression for the anomalous dimension $\eta_T$
\be\label{eq:RG_t}
d t/d \ln y= -t^2\zeta f_1(
\zeta)\;, \quad \eta_T=-
2t\zeta f_0(\zeta) \;.
\ee
The beta function for $t$ is negative, indicating that the system flows to a perfect conductor IR fixed point. In the notation of the main text, we have $f_t(\zeta)\equiv f_1(\zeta)$.

\subsection{Scale-dependent correction to the bosonic propagator Fig.~\ref{fig:leading_N_diag}(f)}
\label{Sec:boson_diagrams}
Another one-loop effect is depicted in Fig.~\ref{fig:leading_N_diag}(f) and describes the corrected bosonic self-energy. First, we find that at zero external momentum and frequency this diagram is IR finite, and thus, can be absorbed into the mass counter-term tuning the system back to criticality. Second, for external $q\neq 0$, $\omega_n\neq 0$ there is a finite small correction to the Landau damping (which can be ignored compared to the tree-level term in \eqref{eq:Pi_z=4_new}), as well as the log divergent correction to the coefficient $c^2$ in front of the bare kinetic term $q^2$. The latter can be used to determine the anomalous dimension $\eta_\phi$ of the bosonic field. To derive this correction, we set $\omega_n=0$ and find
% $\Pi_0(q)=\Pi^{(1)}_0(q)+\Pi^{(2)}_0(q)$ where 
\be 
\Pi_0(q) =4\pi v\gamma \lambda^2 T^2\sum\limits_{\substack{ n_2\\ m_2\leq n_2}}\int_k \Big(   \mathcal{D}_{m_2-n_2,m_4}(k+q)g^{n_2}_{m_2 m_2}(k) -\pi v\gamma T \sum\limits_{ m_4\leq n_2} g_{m_2,m_4}^{n_2}(k+q)g_{m_4,m_2}^{n_2}(k)\Big) \;,
\ee
and $g_{m_2,m_4}^{n_2}(k)\equiv D^\phi_{n_2}(k)\mathcal{D}_{m_2,n_2-m_2}(k)\mathcal{D}_{m_4,n_2-m_4}(k)$. After some algebra, we obtain
\be \label{eq:Pi_z=4}
\Pi_0(q) =-  t\zeta^2\left(f_2(\zeta)-\frac{\zeta}{2}f_3(\zeta) \right) q^2 \ln y\;+ (\text{finite}),
\ee
with the positive monotonic functions
\be
\begin{aligned}
f_2(\zeta)&=\int\limits_0^{+\infty}dy\frac{K_5(y)}{(y+\zeta \mathcal{F}_{1/2}(y))}\approx\begin{cases}
\frac{1}{3}\ln\left(\frac{1}{\zeta}\right),\quad \zeta\ll 1\;\\
0.0962 \frac{1}{\zeta},\quad \zeta \gg 1
\end{cases}\;,\\
f_3(\zeta) &= \int\limits_0^{+\infty}\frac{dy}{\left(y+\zeta\mathcal{F}_{1/2}(y)\right)^2} \left\{\frac{L_2^2(y)}{(y+\zeta\mathcal{F}_{1/2}(y))}\left[1-\zeta K_3(y)-\frac{2y(1-\zeta L_2(y))^2}{(y+\zeta\mathcal{F}_{1/2}(y))} \right]+2 K_4(y)L_2(y) \right.\\
&-\left.2 y L_3^2(y)+\frac{4y L_3(y) L_2(y)}{(y+\zeta\mathcal{F}_{1/2}(y))}\left(\zeta L_2(y)-1\right) \right\}  \approx\frac{1}{\zeta^2}\times\begin{cases}
\theta^2/6,\quad \zeta\ll 1\;\\
 0.0957,\quad \zeta \gg 1
\end{cases}
\end{aligned}
\ee
with
$\theta = \left(\frac{\pi}{2}-1\right)/\left(2-\sqrt{2}\operatorname{arccoth}(\sqrt{2})\right)\approx 0.757$. Here we also defined $K_n(y)=L_{n-1}(y)-2 y L_n(y)$ and $L_{n+1}(y)=\frac{(-1)^{n}}{n!}\partial^{n}_y\mathcal{F}_{1/2}(y)$. The logarithmic divergence in \eqref{eq:Pi_z=4} should be cancelled by the bosonic anomalous dimension 
\be
\eta_{\phi}=-  \frac{1}{2}t\zeta^2\left(f_2(\zeta)-\frac{\zeta}{2}f_3(\zeta) \right),
\ee
which is positive for small $\zeta \ll 1$ and negative for large $\zeta \gg 1$. The very fact that $\eta_\phi$ is not zero is due to the structure of the matrix large-$N$ limit. Indeed, it is known that in the case of the Coulomb interaction (which corresponds to a `vector-type' large-$N$ limit with $N$ fermionic flavours coupled to a single bosonic mode) there is a cancellation between different diagrams resulting in the absence of log corrections to the bare propagator $1/q$. In our situation, such counter-diagrams have different $N$-factors, and thus, do not compensate each other.

\subsection{Scale-dependent correction to the interaction vertex
Fig.~\ref{fig:leading_N_diag}(e)}
Finally, we turn to the vertex correction for the Yukawa coupling $\lambda$. This coupling is present in the quadratic terms in the action, and hence, the RG flows can be computed from
the two-point mixed $W-\phi$ correlator alone. The general form of this diagram is given by 
\be\label{eq:lambda}
    \frac{\delta \lambda }{\lambda} = -\frac{4\pi (v \gamma)^2}{\nu} \;T^2\hspace{-1em}\sum\limits_{\substack{m_2,m_4\\m_2\geq n_2+m_4}}\hspace{-0.5em}\int_k D^\phi_{n_2-m_2}(k+q)D^\phi_{n_1-m_2}(k)\mathcal{D}_{m_4+n_2-m_2,m_2}(k+q)  \mathcal{D}_{n_1,m_2}(k)\mathcal{D}_{m_4+n_2-m_2-n,m_4}(k)\;,
\ee
where all external frequencies $n,n_1,n_2$ and momenta $q$ can be set to zero. After some algebra we obtain
\be
\frac{\delta \lambda}{\lambda} = -\frac{1}{2}  t \zeta^2 f_4(\zeta) \ln y, \;\quad f_4(\zeta)=\int\limits_0^{+\infty}dy\frac{L_2(y)}{(y+1)(y+\zeta \mathcal{F}_{1/2}(y))^2}\approx\begin{cases}
\frac{\theta}{\zeta},\quad \zeta\ll 1\;\\
 \frac{1}{2\zeta^2}\ln\left(\zeta\right),\quad \zeta \gg 1
\end{cases}
\ee
where $f_4(\zeta)$ is positive and monotonic. Now we can write 
\be 
\tilde{\lambda}=(\lambda+\delta \lambda)e^{(\eta_w+\eta_\phi+\eta_T/2)\ln y}= \lambda + \lambda\left(\eta_\phi -t\zeta^2 f_4(\zeta)/2 \right)\ln y\;. 
\ee
By differentiating this expression with respect to $\ln y$, we find the following RG equation for the Yukawa coupling
\be\label{eq:RG_lambda}
\frac{d\ln\lambda}{d\ln y}=-\frac{1}{2}t\zeta  \left(\zeta f_2(\zeta)-\frac{\zeta^2}{2}f_3(\zeta)+\zeta f_4(\zeta)\right)\;.
\ee
The corresponding beta function is always negative, indicating the marginal irrelevancy of $\lambda$. Since all the function $f_i(\zeta)$ depend only on the auxiliary quantity $\zeta$, the closed set of RG equation can be naturally represented in the space $(t,\zeta)$, rather than $(t,\lambda)$. By using \eqref{eq:RG_t} and \eqref{eq:RG_lambda}, it is straightforward to obtain the following RG equation
\be\label{eq:RG_equations1}
% \frac{dt}{d\ln y} &= - t^2 \zeta f_t(\zeta)\;,\quad
% \frac{d\zeta}{ d\ln y} = - t \zeta^2 f_\zeta (\zeta) \;,\\
% f_t(\zeta)&\equiv f_1(\zeta)\;, \quad 
\frac{d\zeta}{ d\ln y} = - t \zeta^2 f_\zeta (\zeta)\;, \quad
f_\zeta(\zeta)\equiv \zeta f_2(\zeta)-\frac{\zeta^2}{2} f_3(\zeta) +\zeta f_4(\zeta)-f_1(\zeta)\;,
\ee
which we discuss in details in the main text. The function $f_\zeta(\zeta)$ has the following asymptotic behavior: $f_\zeta\approx 0.048$ at $\zeta\gg 1$ and $f_\zeta\approx \Delta-1/2$ at $\zeta\ll 1$ where $\Delta= \theta(1-\theta/12)\approx 0.71$. The RG equations \eqref{eq:RG_t} and \eqref{eq:RG_equations1} represent the main result of this section.
\subsection{One-loop correction to the free energy: specific heat}
\begin{figure}[h!]
 \center{\includegraphics[width=0.5\linewidth]{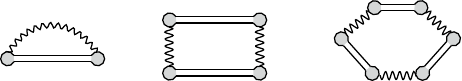}}
\caption{Representative one-loop corrections to the free energy.}
\label{fig:free-energy}
\end{figure}

The temperature dependence of the specific heat can be extracted from interaction corrections to the free energy. The 'classical' contribution yields $c_v\sim T^{2/z}$ with $z=4$, and the representative diagrams contributing to the leading order are depicted in Fig.~\ref{fig:free-energy}. Note that the 'Hartree-type' diagrams are absent due to the last identity in \eqref{eq:large_N_F}. 

After summing up the infinite set of diagrams, we obtain the correction to the free energy density
\be 
\delta \mathcal{F} = -T\sum_n \int_q \ln\left[1+\zeta \frac{\beta_4|\omega_n|^{1/2}}{Dq^2}\mathcal{F}_{1/2}\left(\frac{Dq^2}{\beta_4 |\omega_n|^{1/2}}\right) \right]=\frac{1}{3}\nu\beta_4\; tf_c(\zeta)T^{3/2}+\mathcal{O}(T^0)\;,
\ee
where $f_c(\zeta)=\int_0^{+\infty}dy \ln\left[1+\zeta \mathcal{F}_{1/2}(y)/y \right]$ with $f_c(\zeta)\approx\mathcal{F}_{1/2}(0) \zeta \ln(1/\zeta)$ for $\zeta\ll 1$. For the specific heat, one finds 
\be
\delta c_v = -T\frac{\partial^2 (\delta\mathcal{F})}{\partial T^2} \approx-\frac{\mathcal{F}_{1/2}(0)}{4}\nu \beta_4 t \zeta \ln\left( \frac{1}{\zeta}\right) T^{1/2}\;,
\ee
which results only in the small correction to the coefficient in front of the tree-level temperature dependence $c_v\sim T^{1/2}$, dominated by $\sim N^2$ bosonic modes with dynamical scaling $z_b=4$. This implies that, contrary to the case of disordered FL, the anomalous dimension $\eta_T$ does not modify the $T$--dependence of the specific heat at $T\ll \Lambda_4$.

\section{Singlet diffuson and the density-density correlation function}
\begin{figure}[h!]
 \center{\includegraphics[width=0.5\linewidth]{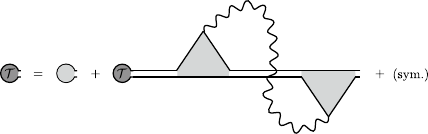}}
\caption{The integral equation describing corrections of the density vertex with $\mathcal{A}=0$.}
\label{fig:charge_vertex}
\end{figure}

In order to compute the density-density correlation function, we need to include the ladder diagrams which we previously ignored in the multiplet $\mathcal{A}\neq 0$ channel in the large-$N$ limit. Specifically, we introduce the density vertex $\mathcal{T}^{\mathcal{A}=0}$ and resum the infinite set of diagrams depicted on Fig.~\ref{fig:charge_vertex}

For simplicity, the density vertex is calculated for zero transferred external momentum, and two frequencies $\omega<0$ and $\omega+\Omega>0$. The integral equation can be formulated in the following form
\begin{multline}
\mathcal{T}^{\mathcal{A}=0}(\omega,\Omega)=1+\frac{\lambda^2}{(4\pi)^2}\int\limits_0^{\Omega}d\epsilon \frac{\mathcal{T}^{\mathcal{A}=0}(\epsilon,\Omega)}{\beta_{4}(|\Omega-\epsilon|^{1/2}+|\epsilon|^{1/2})} \int  \frac{dy}{y+v\gamma D\beta_{4}^{-1}|\omega+\epsilon|^{1/2}\mathcal{F}_{1/2}\left(y/\beta_{4}|\omega+\epsilon|^{1/2}\right)} \times \\
\times\left(\frac{1}{y+\beta_{4}(|\Omega+\omega|^{1/2}+|\epsilon|^{1/2})}+\frac{1}{y+\beta_{4}(|\Omega-\epsilon|^{1/2}+|\omega|^{1/2})} \right)\;.
\end{multline}
The internal integral over $y=Dq^2$ is dominated by the kinematic region where $y>\max \{\beta_{4}|\omega+\epsilon|^{1/2}, \; \beta_{4} (|\Omega+\omega|^{1/2}+|\epsilon|^{1/2}) \}$ for the first term (or $y>\max \{\beta_{4}|\omega+\epsilon|^{1/2}, \; \beta_{4} (|\Omega-\epsilon|^{1/2}+|\omega|^{1/2}) \}$ for the second term). At the saddle point level, each of these two terms produces the factor $(\pi/2)(v\gamma D)^{-1/2} |\omega+\epsilon|^{-1/2}$. Additionally, the pre-factor is simply given by $\lambda^2/((4\pi)^2\beta_{4})=(v\gamma D)^{1/2}/(2\pi)$. Thus, combining everything together, we obtain
\be
\mathcal{T}^{\mathcal{A}=0}(\omega, \Omega)=1+\frac{1}{2}\int\limits_{0}^{\Omega}\frac{d\epsilon}{|\epsilon+\omega|^{1/2}}\frac{\mathcal{T}^{\mathcal{A}=0}(\epsilon,\Omega)}{\left(\Omega/(N\beta_{4})+\epsilon^{1/2}+(\Omega-\epsilon)^{1/2}\right)} \;,
\ee
where we also re-introduced the bare frequency dependence $\Omega/N$ of the diffuson at energy scales higher than $\Lambda_4$. One can rescale $\epsilon=\Omega t$ and note that $\mathcal{T}^{\mathcal{A}=0}$ is a function of two dimensionless parameters: $x=|\omega|/\Omega\;<1$ and $\delta = \Omega^{1/2}/(N\beta_{4})$. The integral equation takes the form

\be\label{eq:density_vertex_int}
\mathcal{T}^{\mathcal{A}=0}(x,\delta)=1+\frac{1}{2}\int\limits_{0}^{1}\frac{dz}{|x-z|^{1/2}}\frac{\mathcal{T}^{\mathcal{A}=0}(z,\delta)}{\left(\delta +z^{1/2}+(1-z)^{1/2}\right)}\;.
\ee
One can easily verify that its solution is given by
\be
\mathcal{T}^{\mathcal{A}=0}(x,\delta)=\delta^{-1}\left(\delta + x^{1/2}+(1-x)^{1/2} \right)\;.
\ee
Now it becomes clear why we had to introduce some UV regularization - the series in \eqref{eq:density_vertex_int} is not geometrical and diverges when $\delta\rightarrow 0$. The appropriate UV completion of the theory is required in order to satisfy the Ward identity: the particle conservation implies that one needs to pick up contributions to the density-density correlator from all energy scales. As a result, for the dynamical part of the density-density correlation function we obtain
\be
\Pi^{\mathcal{A}=0}(\Omega,Q=0) =- \delta\;\frac{dn}{d\mu}\int_0^{1}dx \frac{\mathcal{T}^{\mathcal{A}=0}(x,\delta)}{((1-x)^{1/2}+x^{1/2})} \simeq -\frac{dn}{d\mu} 
\ee
where $dn/d\mu$ is the thermodynamic density of states. Note that this contribution exactly cancels the static part of the density-density correlation function, as expected from the Ward identity. In addition, our derivation can be easily modified to account for the finite external momentum transfer $Q\neq 0$, which leads to a diffusion pole with the dynamical scaling $z=2$ for a singlet mode. However, we note that the logarithmic corrections to the saddle point, discussed in the previous sections in relation to the multiplet diffusons, inevitably lead to renormalization of the diffusion coefficient according to the RG equations (7) of the main text.

\end{document}